\documentclass[12pt]{article} 
 
\begin{document} 
 
\title{A new approach to the parametrization of the Cabibbo-Kobayashi-Maskawa matrix} 
\author{V. Gupta \\ 
Departamento de F\'\i sica Aplicada,CINVESTAV-Unidad M\'erida,\\ 
A.P. 73 Cordemex 97310 M\'erida, Yucat\'an, M\'exico} 
\maketitle 
 
\begin{abstract} 
The CKM-matrix V is written as a linear combination of the unit matrix I and 
a matrix U which causes intergenerational-mixing. It is shown that such a V 
results from a class of quark-mass matrices. The matrix U has to be 
hermitian and unitary and therefore can depend at most on 4 real parameters. 
The available data on the CKM-matrix including CP-violation can be 
reproduced by $V=(I+iU)/\sqrt{2}$. This is also true for the special case 
when U depends on \textit{only 2 real parameters}. There is no CP-violating 
phase in this parametrization. Also, for such a V the invariant phase $\Phi 
\equiv \phi _{12}+\phi _{23}-\phi _{13}$, satisfies a criterion suggested 
for `maximal' CP-violation. 
\end{abstract}

It is more than twenty-five years since the first explicit parametrization 
for the six quark case was given ~\cite{A} for the so called 
Cabibbo-Kobayashi-Maskawa (CKM) matrix. Since then many diferent 
parametrizations have been suggested ~\cite{B,C}. In this note, we wish to 
suggest a new approach to parametrizing the unitary CKM matrix V. For this 
purpose, we write V as a linear combination of the unit matrix $I$ and 
another matrix $U$, so that 
 
\begin{equation} 
V(\theta )=\cos \theta I+i\sin \theta \,U  \label{1} 
\end{equation} 
 
It is clear that for V to be unitary, U has to be both hermitian and 
unitary. Here $\theta $ is a parameter which will be fixed later. In Eq. 
(1), for the first term the physical (or the quark mass-eigenstate) and the 
gauge bases are the same. The second term, through U, represents the 
difference in the two bases. It also causes inter-generational mixing and 
makes it possible for V to give CP-violating processes. The break-up of V in 
two parts makes it possible to have a simple parametrization. We now show 
that knowing $V(\theta)$ allows us to construct the quark-mass matrices in 
terms of the parameters of V and the quark-masses. 
 
\textit{Form of the quark-mass matrices.} In the gauge-basis, the part of 
the standard model Lagrangian relevant for us can be written as 
 
\begin{equation} 
\mathcal{L}=\bar{q}_{uL}^{\prime }M_{u}q_{uR}^{\prime }+\bar{q}_{dL}^{\prime 
}M_{d}q_{dR}^{\prime }+\frac{g}{\sqrt{2}}\bar{q}_{uL}^{\prime }\gamma _{\mu 
}q_{dL}^{\prime }W_{\mu }^{+}+H.c.  \label{2} 
\end{equation} 
 
\noindent where ${q} ^{\prime}_{u}= (u^{\prime},c^{\prime},t^{\prime})$ and $%
{q} ^{\prime}_{d}= (d^{\prime},s^{\prime},b^{\prime})$. By suitable 
redefinition of the right-handed quark fields one can make the quark-mass 
matrices $M_u$ and $M_d$ hermitian. Let the diagonal forms of the hermitian $%
M_u$ and $M_d$ be given by 
 
\begin{equation} 
\hat{M}_{u}=V_{u}^{\dagger }M_{u}V_{u},\hspace{1cm}\hat{M}%
_{d}=V_{d}^{\dagger }M_{d}V_{d}.  \label{3} 
\end{equation} 
 
\noindent In the physical basis, defined by $q_{\alpha }=V_{\alpha 
}^{\dagger }q_{\alpha }^{\prime }$ ($\alpha =u$ or $d$), one has  
\begin{equation} 
\mathcal{L}=\sum_{\alpha =u,d}\bar{q}_{\alpha L}\hat{M}_{\alpha }q_{\alpha 
R}+\frac{g}{\sqrt{2}}\bar{q}_{uL}\gamma _{\mu }Vq_{dL}W_{\mu }^{+}+H.c. 
\label{4} 
\end{equation} 
 
\noindent where 
 
\begin{equation} 
V=V_{u}^{\dagger }V_{d}  \label{5} 
\end{equation} 
is the CKM-matrix. 
 
For a V given by Eq.(1), one can easily find $V_{u}$ and $V_{d}$ which 
satisfy Eq. (5) In general,  
\begin{equation} 
V_{u}=V(\theta _{u})=\cos \theta _{u}I-i\sin \theta _{u}U  \label{6} 
\end{equation} 
\begin{equation} 
V_{d}=V(\theta _{d})=\cos \theta _{d}I+i\sin \theta _{d}U  \label{7} 
\end{equation} 
 
\noindent will give $V(\theta)$ provided $\theta_u + \theta_d = \theta$. 
This is so since $V( \theta _1) V( \theta _2) = V( \theta_1 + \theta_2)$ 
because $U=U^{\dagger}$ and $U^2=I$. 
 
Given these $V_u$ and $V_d$, Eq.(3) then \textit{determines} $M_u$ and $M_d$ 
in terms of the quark masses and the experimentally accessible parameters of 
the CKM-matrix. More formally, this means that in the spectral decomposition 
of $M_u (M_d)$ the projectors depend only on the parameters in $V(\theta)$ 
and $\theta _u( \theta _d)$. There is a freedom in the choice of the values $%
\theta_u$ and $\theta_d $ as only their sum $\theta _u + \theta _d = \theta$ 
is determined from knowing $V( \theta)$. 
 
It is clear that our form of $V(\theta)$ provides an explicit solution for a 
class of quark mass matrices. 
 
\textit{Form of U in the standard model.} To determine the general form of 
the hermitian and unitary $3\times 3$ matrix $U$ we start with a general 
hermitian matrix  
\begin{equation} 
U=\left(  
\begin{array}{ccc} 
u_{1} & \alpha ^{*} & \beta ^{*} \\  
\alpha & u_{2} & \gamma ^{*} \\  
\beta & \gamma & u_{3} 
\end{array} 
\right)  \label{8} 
\end{equation} 
where $u_{i}\ (i=1,2,3)$ are real and $\alpha ,\beta $ and $\gamma $ are 
complex numbers. Requiring $U$ to be unitary as well implies that $U^{2}=I.$ 
Explicitly this gives  
\begin{equation} 
u_{1}^{2}+\left| \alpha \right| ^{2}+\left| \beta \right| ^{2}=1,  \label{9} 
\end{equation} 
\begin{equation} 
u_{2}^{2}+\left| \alpha \right| ^{2}+\left| \gamma \right| ^{2}=1, 
\label{10} 
\end{equation} 
\begin{equation} 
u_{3}^{2}+\left| \beta \right| ^{2}+\left| \gamma \right| ^{2}=1;  \label{11} 
\end{equation} 
and  
\begin{equation} 
\left| \alpha \right| \left( u_{1}+u_{2}\right) +\left| \beta \gamma \right| 
\exp (i\phi )=0,  \label{12} 
\end{equation} 
\begin{equation} 
\left| \beta \right| \left( u_{1}+u_{3}\right) +\left| \alpha \gamma \right| 
\exp (-i\phi )=0,  \label{13} 
\end{equation} 
\begin{equation} 
\left| \gamma \right| \left( u_{2}+u_{3}\right) +\left| \alpha \beta \right| 
\exp (i\phi )=0.  \label{14} 
\end{equation} 
Here $\phi \equiv \phi _{\alpha }-\phi _{\beta }+\phi _{\gamma }$ while $%
\phi _{\alpha },$ $\phi _{\beta }$ and $\phi _{\gamma }$ are the phases of $%
\alpha ,$ $\beta $ and $\gamma .$ Eqs. (\ref{12}-\ref{14}) immediatly imply 
that $\sin \phi =0$ or $\phi =0$ or $\pi .$ The resulting $U$ in the two 
cases differ by an overall sign \cite{D}. For definiteness we consider the 
case $\phi =0.$ Eqs. (\ref{12}-\ref{14}) determine the diagonal elements in 
terms of $\left| \alpha \right| ,$ $\left| \beta \right| $ and $\left| 
\gamma \right| $ and substituting these in Eqs. (\ref{9}-\ref{11}) gives the 
constraint  
\begin{equation} 
\left| \frac{\alpha \beta }{\gamma }\right| +\left| \frac{\alpha \gamma }{%
\beta }\right| +\left| \frac{\beta \gamma }{\alpha }\right| =2.  \label{15} 
\end{equation} 
Using this one has  
\begin{equation} 
\begin{array}{ccc} 
u_{1}=\left| \frac{\alpha \beta }{\gamma }\right| -1, & u_{2}=\left| \frac{%
\alpha \gamma }{\beta }\right| -1 & \mbox{and }u_{3}=\left| \frac{\beta 
\gamma }{\alpha }\right| -1. 
\end{array} 
\label{16} 
\end{equation} 
For a more convenient form of $U,$ we put 
\begin{equation} 
\begin{array}{ccc} 
\alpha =-2bc^{*}, & \beta =-2ac, & \mbox{and }\gamma =-2a^{*}b. 
\end{array} 
\label{17} 
\end{equation} 
Since, $\phi _{\alpha }=(\phi _{b}-\phi _{c})+\pi $ etc., the condition $%
\phi =0$ translates into  
\begin{equation} 
\phi _{a}-\phi _{b}+\phi _{c}=\frac{\pi }{2},  \label{18} 
\end{equation} 
where $\phi _{a},$ $\phi _{b}$ and $\phi _{c}$ are the phases of the complex 
numbers $a,$ $b$ and $c.$ The constraint of Eq. (\ref{15}) becomes  
\begin{equation} 
\left| a\right| ^{2}+\left| b\right| ^{2}+\left| c\right| ^{2}=1.  \label{19} 
\end{equation} 
 
The general expression of the hermitian and unitary $U$ in terms of $a,$ $b$ 
and $c$ is  
\begin{equation} 
U=I-2\left(  
\begin{array}{ccc} 
|a|^{2}+|b|^{2} & b^{*}c & a^{*}c^{*} \\  
bc^{*} & |a|^{2}+|c|^{2} & ab^{*} \\  
ac & a^{*}b & |b|^{2}+|c|^{2} 
\end{array} 
\right)  \label{20} 
\end{equation} 
Given the two constraints in Eq. (\ref{18}) and Eq. (\ref{19}), we note that 
a general hermitian and unitary $3\times 3$ matrix depends on at most four 
real parameters. This is the form of $U$ we will use \cite{D}. 
 
The Jarslskog invariant ~\cite{E} for $U$, viz. $%
J(U)=Im(U_{11}U_{22}U_{12}^{*}U_{21}^{*})=0.$ However, the $V(\theta )$ in 
Eq. (1) does give CP-violation, since 
 
\begin{equation} 
J(V(\theta ))=8\cos \theta \sin ^{3}\theta |abc|^{2}=\cos \theta 
|V_{12}||V_{13}||V_{23}|  \label{21} 
\end{equation} 
In our case, there is no `CP-violating phase' which governs the finitess of $%
J$. One of the off-diagonal elements of $V(\theta )$ has to be zero for $J$ 
to vanish. Note, that J is just given in terms of $|V_{ij}|(i\neq j)$ unlike 
usual parametrizations ~\cite{C}. It is interesting to note, that even when $%
a$,$b$,$c$ are pure imaginary ~\cite{F} so that $V(\theta )$ depends on only 
3 real parameters, $J(V(\theta ))$ is non-zero. In this case, $U$ becomes 
real and symmetric and the only complex number in $V(\theta )$ is $i$ in 
Eq.(1) ! 
 
Since $U$ is hermitian it requires that $|V_{ij}|=|V_{ji}|$ for $V(\theta )$ 
in Eq. (\ref{1}). The experimentally determined CKM-matrix $V_{EX}$ given by 
the Particle Data Group \cite{C} 
 
\begin{equation} 
V_{EX}=\left(  
\begin{array}{ccc} 
0.9745-0.9760 & 0.2170-0.2240 & 0.0018-0.0045 \\  
0.2170-0.2240 & 0.9737-0.9753 & 0.0360-0.0420 \\  
0.0040-0.0130 & 0.0350-0.0420 & 0.9991-0.9994 
\end{array} 
\right)  \label{22} 
\end{equation} 
 
\noindent The entries correspond to the ranges for the moduli of the matrix 
elements. It is clear that $|V_{12}|=|V_{21}|$ and $|V_{23}|=|V_{32}|$ are 
satisfied for the whole range, while the equality $|V_{13}|=|V_{31}|$ is 
suggested by the data. Given the fact that $|V_{13}|$ and $|V_{31}|$ are the 
hardest to determine experimentally, it is possible they might turn out to 
be equal. We adopt a common numerical value viz. $|V_{13}|=|V_{31}|=0.005825%
\pm 0.002925.$ This numerical value is obtained by first converting the 
range of values in $V_{EX}$ into a central value with errors, so that $%
|V_{13}|=0.00315\pm 0.00135$ and $|V_{31}|=0.0085\pm 0.0045.$ The average of 
these two gives the common numerical value above. Ranges for other moduli 
also are converted into a central value with errors. 
 
To confront $V(\theta )$ with experiment we need to specify $\theta $. A 
physically appealing choice is to give equal weight to the generation mixing 
term $(U)$ and the generation diagonal term $(I)$ in $V(\theta )$, so that $%
\theta =\pi /4$ and the CKM-matrix 
 
\begin{equation} 
V(\pi /4)=\frac{1}{\sqrt{2}}(I+iU).  \label{23} 
\end{equation} 
 
\noindent We use this for numerical work. 
 
\textit{Numerical results} Experimentaly, $|V_{12}|$ and $|V_{23}|$ are well 
determined. We take their average (or central) value in the range given in 
Eq. (\ref{22}) as inputs; that is, $|V_{12}|=|V_{21}|=0.2205$ and $%
|V_{23}|=|V_{32}|=0.039$. Given these, one has 
 
\begin{equation} 
|a|=|V_{23}|/(2\sin \theta |b|),  \label{24} 
\end{equation} 
 
\begin{equation} 
|c|=|V_{12}|/(2\sin \theta |b|).  \label{25} 
\end{equation} 
 
The constraint Eq. (\ref{19}), gives a quadratic equation for $|b|^{2}$ with 
the solutions,  
\begin{equation} 
|b|^{2}=\frac{1}{2}\left[ 1\pm \sqrt{1-(|V_{12}|^{2}+|V_{23}|^{2})\csc 
^{2}\theta }\right] .  \label{26} 
\end{equation} 
 
\noindent Note, for real $|b|^{2}$, above input implies $\sin ^{2}(\theta 
)\geq 0.05014$ or $\theta \geq 12.94^{\circ }$. Since, $%
|V_{12}|>|V_{23}|>|V_{13}|$ it is clear we need the positive sign in Eq. (%
\ref{26}) so that $|b|>|c|>|a|$. For $\theta =\pi /4$, Eqs. (\ref{24}-\ref 
{26}) yield, 
 
\begin{equation} 
|a| = 0.02794, \hspace{1cm} |b|= 0.98705, \hspace{1cm}|c|= 0.15796 .  
\end{equation} 
 
The values of the $|V_{ij}|$ for $V(\pi /4)$ in Eq. (\ref{23}) are given in 
Table I. The values in the table should be compared with the average values 
of $|V_{ij}|$ obtained from $V_{EX}$. For example, average of $V_{11}$ from 
Eq. (\ref{22}) is $\frac{1}{2}(0.9745+0.9760)=0.97525$. This is given as $%
0.97525\pm 0.00075$. The `error' indicates the range for $|V_{11}|$. The 
experimental $|V_{ij}|$ are given in column 2, while the calculated values 
are given in column 3. The agreement is quite satisfactory suggesting that a 
CKM-matrix with$|V_{ij}|=|V_{ji}|$ may fit the data. We did not attempt a 
best fit in view of our assumption $|V_{13}|=|V_{31}|$ . 
 
The value of $J$ for $V_{EX}$ and $V(\pi /4)$ are also given in the Table. $%
J(V_{EX})$ was calculated using the formula ~\cite{G}  
\begin{equation} 
J^{2}=|V_{11}V_{22}V_{12}V_{21}|^{2}-\frac{1}{4}[%
1-|V_{11}|^{2}-|V_{22}|^{2}-|V_{12}|^{2}-|V_{21}|^{2}+|V_{11}V_{22}|^{2}+|V_{12}V_{21}|^{2}%
]^{2}  \label{27} 
\end{equation} 
\noindent with the central values of $|V_{ij}|$, $i=1,2$ and since these 
four are best measured. The value $J(V(\pi /4))$ was calculated using Eq. (%
\ref{21}) and is about $3-4$ times smaller. This is reasonable considering 
the slight differences in values of $|V_{i,j}|$ $i=1,2$ in the two cases and 
also since there is a strong numerical cancellation between the two terms on 
the r.h.s of Eq. (\ref{27}). 
 
It is important to note that calculated values require only the knowledge of  
$|a|$, $|b|$ and $|c|$. Thus, the numerical results are valid even when $a$,  
$b$ and $c$ are pure imaginary ~\cite{F} and $V(\pi /4)$ depends on only 2 
real parameters ~\cite{H}. 
 
\textit{Concluding remarks} Apart, form providing a good numerical fit with 
4 or possibly 2 parameters, the CKM-matrix $V(\pi /4)$ has an interesting 
feature connected with a criterion ~\cite{I} for `maximal' CP-violation. 
 
It was noted ~\cite{I} that physically the relevant phase for CP-violation 
in the CKM-matrix $V$ is $\Phi =\phi _{12}+\phi _{23}-\phi _{13}$, where $%
\phi _{ij}$ is the phase of the matrix element $V_{ij}$. The reason for this 
is because $\Phi $ is invariant under re-phasing transformations of $V$. So, 
a value of $\Phi \equiv |\pi /2|$ was suggested as corresponding to 
`maximal' CP-violation. This is so in our case because of the constraint in 
Eq. (\ref{18}) since $\Phi =2(\phi _{a}+\phi _{c}-\phi _{b})-\pi /2$. So, $%
\cos \Phi =0$ for $V(\pi /4)$. Note that, $\Phi =\pi /2$ is automatic when $%
a $, $b$ and $c$ are pure imaginary ~\cite{F} and in that case $V(\pi /4)$ 
depends on only 2 real parameters. 
 
It is remarkable that $V(\pi /4)$ with only 2 real parameters fits the 
available data. This may be because only the absolute values $|V_{ij}|$ are 
known at present. Future information on the full $V_{ij}$ will tell us if 
the relations ~\cite{J} implied by the two parameter parametrization given 
here are viable or the more general four parameter parametrization would be 
needed. It would be very interesting if the symmetry relations $%
|V_{ij}|=|V_{ji}|(i\ne j)$ are confirmed experimentally. 
 
Acknowledgments. I am grateful to Dr. A.O. Bouzas for critical comments and 
discussion. I am thankful to Vicente Antonio Perez, Elena Salazar and Yuri 
Nahmad for their help in preparing the manuscript.

\pagebreak 
 
\[ 
\begin{tabular}{||c|c|c||} 
\hline\hline 
Quantity & Experiment & Theory \\ \hline 
$|V_{12}|=V_{21}$ & $0.2205\pm 0.0035$ & $0.2205$ (input) \\ \hline 
$|V_{23}|=V_{32}$ & $0.0390\pm 0.0030$ & $0.039$ (input) \\ \hline 
$|V_{13}|=V_{31}$ & $0.005825\pm 0.002925$ & $0.00624$ \\ \hline 
$V_{11}$ & $0.97525\pm 0.00075$ & $0.975367$ \\ \hline 
$V_{22}$ & $0.9745\pm 0.0008$ & $0.974607$ \\ \hline 
$V_{33}$ & $0.99925\pm 0.00015$ & $0.99922$ \\ \hline 
$J$ & $1.414\times 10^{-4}$ & $3.795\times 10^{-5}$ \\ \hline\hline 
\end{tabular} 
\] 
\centerline{\parbox{4 true in}{\footnotesize {\bf Table I.} Numerical 
values of the moduli of the matrix elements of $V (\theta)$ for  
$\theta = \pi /4$. Experimental values are average values obtained from 
$V_{EX}$ in Eq. (13). The `errors' reflect the large of values for 
$|V_{ij}|$. 
Note, since $|V_{13}|=0.00315 \pm 0.00135 $ and $|V_{31}|= 0.0085 \pm  
0.0045$, we grote the average of these in the Table. $J$ is the Jarslskog 
invariant (see text)}} 
 

\begin{thebibliography}{99}
\bibitem{D}  In case of the choice $\phi =\pi ,$ $u_{i}$ have the opposite 
sign, so that $u_{1}=1-\left| \frac{\alpha \beta }{\gamma }\right| $ etc. 
Then puting $\alpha =+2bc^{*}$ etc. gives $\phi _{a}-\phi _{b}+\phi _{c}=-%
\frac{\pi }{2}$ and the resulting matrix is just -1 times $U$ given in Eq. (%
\ref{20}). Note in either case, $\cos \left( \phi _{a}-\phi _{b}+\phi 
_{c}\right) =0.$
\bibitem{F}  Note that for pure imaginary a, b and c, one has $\phi 
_{a}=\phi _{b}=\phi _{c}=\pi /2$. So Eq. (\ref{18}) is automatically 
satisfied. In this case, $U$ and $V(\pi /4)$ depend on only two real 
parameters.
\bibitem{J}  For example $|V_{ij}|=|V_{ji}|(i\ne j)$ . Also phases $\phi 
_{ij}$ are fixed and other relations betwen absolute magnitudes exist. For $%
V(\pi /4)$, an interesting relation (implied by Eq.(\ref{19}) and Eq.(18)) 
is $2J=\sqrt{2}%
|V_{12}V_{13}V_{23}|=|V_{12}V_{23}|^{2}+|V_{12}V_{13}|^{2}+|V_{23}V_{13}|^{2} 
$
\end{thebibliography}
\end{document}